
\baselineskip=20pt
\parindent 20pt
\settabs 4 \columns
\parskip=10pt
\hfuzz=2pt   
\def\line{\hbox to \hsize}
\def\frac #1#2{{#1\over #2}}


\line{\hfil }
\line{\hfil IP/BBSR/95-54}
\line{\hfil April 1995}
\vskip 1cm
\centerline{\bf ``CLASSICAL'' VORTEX NUCLEATION IN SUPERFLOW}
\centerline{\bf    THROUGH SMALL ORIFICES  }
\vskip 1cm

\centerline{Michael Stone}
\centerline{\it Department of Physics}
\centerline{\it University of Illinois at Urbana Champaign}
\centerline{\it 1110 W. Green St.}
\centerline{\it Urbana, IL 61801}
\centerline{USA}
\centerline{\&}
\centerline{Ajit M. Srivastava}
\centerline{\it Institute of Physics}
\centerline{\it Sachivalaya Marg }
\centerline{\it Bhubaneswar 751005}
\centerline{\it INDIA}
\line{\hfil}

\line{\bf Abstract\hfil}

We report on numerical solutions of the Gross-Pitaevskii equation for
two-dimensional flow of a superfluid condensate through a small
orifice. Above a critical velocity of about $30\%$ of the speed of
sound, cavitation occurs in the throat of the orifice. The cavitating
bubbles form the cores of singly quantized vortices which detach
from the boundary  and are convected downstream.

\vfil

\eject

\line{\bf 1. Introduction\hfil}

Boundary layer separation  is a familiar phenomenon in classical fluid
dynamics [1]. It is invariably associated with the appearance of
vorticity in the bulk of the fluid. The  main flow and the relatively
stagnant fluid in the lee of the bounding body may be in irrotational
motion but they are separated by  the detached boundary layer, a  sheet
of fluid containing vorticity generated by viscous forces while it was
in contact with the body. For example, when a classical fluid is
forced  through a small opening the flow separates close to the
narrowest point  of the orifice and forms a jet penetrating
into the ambient fluid.  This jet is bounded by a tube
of vorticity which confines the flow  much as a solenoid confines the
magnetic field it produces.  When the opening has a sharp salient edge
the boundary layer separation takes place at arbitrarily low flow
velocities. If the edge is rounded there will be some critical velocity
below which the flow remains attached and potential.

Similar regimes occur when a superfluid is forced through a small
opening.  At low velocities the superflow is potential and
dissipationless. Liquid can pass through the orifice without the need
for a pressure difference between the two sides. Above a critical flow
rate a pressure head is needed to maintain the flow. At
sufficiently high flow rate even a superfluid  behaves  as a classical
liquid, and in this regime we can be confident that a collimated jet
has formed.  Because vorticity is quantized in a superfluid, the flow
separation requires the shedding of discrete vortices from the
boundary. These vortices will approximate the classical vortex sheet
bounding the  the  jet, just as the individual windings approximate an
ideal solenoid.

In recent years resonators containing micron sized orifices have been
constructed in which one can monitor the dissipation produced by the
shedding of a single  vortex [2,3]. At very low temperatures the
critical velocities in the orifice become independent of temperature,
and this temperature independence is cited as evidence that  the
vortices  are being created  by quantum mechanical tunneling [4,5].
This inference is probably correct, but it is worth appreciating that
thermal or quantum nucleation are not the only alternatives for
creating a vortex.  Vortices may also be shed from a boundary via
cavitation.  An accurate estimate of the critical velocity for this
``classical'' vortex creation mechanism is needed before one can state
with confidence that one is seeing  macroscopic quantum tunneling
events.

This present paper provides a rough estimate of this critical velocity.
It reports numerical solutions of the time-dependent Gross-Pitaevskii
(G-P) equation [6,7] for flow through (two dimensional) orifices.
These solutions demonstrate that when the velocity near the classical
flow separation point exceeds about $30\%$ of the speed of sound, low
pressure in the throat of the orifice causes cavitation.  Pairs of
bubbles are pulled  off the boundary, the pair forming the core of what
in three dimensions would be a singly-quantized smoke-ring-like
vortex.

It is surprising that this classical  cavitation  effect was not
discovered long ago. It is not described in any of the standard
references on the G-P equation but, while we were performing the
calculations reported here, our attention was drawn to the recent work
of Frisch, Pomeau and Rica who observed similar vortex shedding
from  the boundary of cylinders immersed in a uniform flow [8,9].
These authors interpret the vortex shedding as a substitute for
shock-wave solutions (which do not occur in the G-P equation) rather
than, as seems more natural to us, an emulation of  boundary layer
separation.  Also they do not exhibit  detailed pictures of the
formation of the vortices.  Consequently, in addition to readvertizing
the phenomenon and estimating the critical velocity, we include some
detailed plots of how the order parameter phase evolves as the vortices
form. In the second section we will review the well-known interpretation of
the real and imaginary parts of the  Gross-Pitaevskii equation as the
equations of fluid mechanics. We  also include a discussion of the
Josephson phase winding between the core of a jet and the ambient
fluid.  In the  third section we will report on the numerical methods and
boundary conditions used to induce the flows. In the  fourth section we
describe the results.

\vskip 40pt
\line {\bf 2) The Gross-Pitaevskii equation\hfil }

The Gross-Pitaevski (G-P) non-linear Schr{\"o}dinger equation [6,7]
provides a simple  model for the dynamics of the condensate in
$^4He$.  Although the local version of this equation does not describe
the roton part of the quasiparticle spectrum [9], it does model the
most essential ingredients: long-range phase coherence coupled with mass and
momentum conservation.

To establish our notation, and to develop some physical insight, we
begin by reviewing  the close relationship between the G-P equation and
classical fluid dynamics. Because it is important to distinguish
between quantum and  classical effects we will include explicit factors
of $\hbar$.  For generality we will also include a coupling to a gauge
field, although the application in this paper is primarily to neutral
superfluid $^4He$.

The Gross-Pitaevskii equation is the non-linear Schr{\"o}dinger
equation  of the form
$$
i\hbar(\partial_t-ieA_0/\hbar)\Psi=-\frac
{\hbar^2}{2m}\sum_{a=1}^{3}(\partial_a-ieA_a/\hbar)^2\Psi +\lambda
(|\Psi|^2-\rho_0)\Psi.  \eqno (1.1)
$$
The sign of $\lambda$ will be
understood to be positive, so the last term in (1.1) represents a
repulsive force between the the $^4He$ atoms.  (This is the opposite
sign from that usually taken when studying the one-dimensional
non-linear Shr{\"o}dinger equation as an example of an integrable
soliton system.) The mass $m$ should be taken to be that  of a helium
atom ($6.647\times 10^{-27}\,{\rm kg}$). The values of $\lambda$,  and
the number density $\rho_0$ are usually chosen to fit the measured
density ($m\rho_0 =145\, {\rm kg\, m}^{-3}$), and speed of sound
($c=\sqrt{\lambda\rho_0/m}=230\,{\rm ms}^{-1}$) in the  fluid.

By using the Madelung transformation [10], (1.1) can be recast
as the equation of motion of a charged compressible
fluid having equilibrium particle-number density $\rho_0$.
We set $\Psi=\sqrt{\rho}e^{i\theta}$  and
define a velocity field ${\bf v}$ in such a way that the number-current,
$$
{\bf j}=\frac {\hbar}{2mi}\left(\Psi^*(\nabla -ie{\bf A}/\hbar)
\Psi-((\nabla+ie{\bf A}/\hbar)\Psi^*)\Psi\right),
\eqno(1.2)
$$
may be written as ${\bf j}=\rho{\bf v}$. This requires
$$
{\bf v}=\frac \hbar m (\nabla\theta-e{\bf A}/\hbar).
\eqno(1.3)
$$

In the absence of vortex singularities in $\Psi$, the  vorticity,
$\omega=\nabla\wedge {\bf v}$, is completely determined by the
gauge field to be
$
\omega=-\frac em \nabla\wedge {\bf A}=-\frac{e{\bf B}}{m},
$
{\it i.e}
$$
m\omega+e{\bf B}=0.
\eqno(1.4)
$$

For neutral superfluids (1.4) implies irrotational motion and hence, at
low velocities, where the effects of compressibility can be ignored,
leads to D'Alembert's paradox (the absence of drag forces).  For
charged superfluids  equation (1.4) is responsible for the Meissner
effect: A penetrating uniform ${\bf B}$ field would require  uniform
vorticity, {\it i.e}, rigid rotation. A rigidly rotating body possesses a
kinetic energy which grows faster than the volume of the system and so
is impossible in the thermodynamic limit.  Alternatively, taking the
curl of the Maxwell equation
$
{\nabla \wedge} {\bf B}=e{\bf j}$ and using  ${\bf j}=\rho {\bf v}_s$
 together with (1.4), implies that
$$
\nabla^2 {\bf B}-\frac {e^2\rho}{ m }{\bf B}=0,
\eqno(1.5)
$$
which leads to flux screening. Note that $\hbar$ does not appear in (1.4)
or in the screening length $(e^2\rho/m)^{-1/2}$.

With the definition (1.3), the imaginary  and real parts of (1.1)
become, respectively, the continuity equation
$$
\partial_t\rho+\nabla\cdot \rho{\bf v}=0,
\eqno(1.6)
$$
and the Euler equation  governing the flow of a barotropic fluid
$$
m(\partial_t {\bf v}+{\bf v}\cdot \nabla {\bf v})=
e({\bf E}+{\bf v}\wedge {\bf B})-\nabla \mu.
\eqno(1.7)
$$
The word {\it barotropic\/}  refers to the
simplifying property that the pressure term $\frac1\rho\nabla P$,
which occurs on the right hand side of the conventional Euler
equation, is here combined into the gradient of a potential
$$
\mu\equiv\lambda(\rho-\rho_0)-\frac {\hbar^2}{2m}\frac{\nabla^2 \rho}{\rho}.
\eqno(1.8)
$$ The potential $\mu$ contains the expected compressibility pressure,
 depending on the deviation from the equilibrium density, together with a
correction depending on gradients of $\rho$. This correction, the  {\it
quantum pressure}, is the only place that $\hbar$ appears in the flow
equations.
It sets the length scale $\xi=\hbar/\sqrt{2m\lambda \rho_0}$
over which the superfluid density heals after being forced to zero by a
boundary or a vortex singularity. With the parameters chosen as
suggested above we find $\xi=.487 \,{\rm \AA}$.  ($\hbar$ also manifests
itself,
of course, in the quantum of circulation, $\kappa=2\pi \hbar/m$. )

The Euler equation (1.7) is   derived by first taking the gradient of
(1.1)  and interpreting the result  as the Bernoulli equation,
$$
m(\partial_t {\bf v} -{\bf v}\wedge \omega)= e({\bf E}+{\bf v}\wedge{\bf
B})-\nabla\left(\frac 12 m {\bf v}^2+\mu\right),
\eqno(1.9)
$$
which is equivalent to (1.7).

In (1.9) a cancellation  of the $m{\bf v}\wedge \omega$ term against
the  $e{\bf v}\wedge{\bf B}$  term is evident upon use of (1.4).  It is
after this cancellation, and so  without reference to either ${\bf B}$
or $\omega$, that the hydrodynamic picture of superconductivity is
conventionally displayed. It seems  preferable to keep both $\omega$
and $B$   in (1.9) and rewrite it as (1.7).  By doing this one can see
that the only difference between  superfluid dynamics and classical
fluid dynamics lies in the constraint (1.4).

In the next section we solve (1.1)  numerically for the case
of a neutral superfluid forced to flow through an orifice. Let us
first consider a crude model of the extreme case  where the downstream
flow forms a tubular jet penetrating into superfluid at rest.

We suppose that there is a pressure head, and correspondingly a
difference  $\mu_1-\mu_0$ in the potential,  between the asymptotic
parts of the reservoirs communicating via the orifice.  This  means
that the phase of the order parameter of the fluid at rest  in the
upstream reservoir, $\theta_1$,  must be falling behind the phase of
the order parameter in of the fluid at rest
in the lower presure container, $\theta_0$, at a rate given by
$$
-\hbar\frac {d\phantom t}{dt}(\theta_1-\theta_0) =\mu_1-\mu_0.
\eqno (1.10)
$$
If the flow is steady, (1.9) implies that $\dot \theta$ is position
independent.  Consequently, as we follow a streamline through the
orifice into the core of the jet, we must have the classical Bernoulli
relation
$$
\frac 12 mv^2+\mu= {\rm const.}
\eqno (1.11)
$$

Now the pressure in the jet and that of the adjacent ambient fluid must
be the same, or the fluid would move sideways. This requires the
value of  $\mu$ in the jet to equal $\mu_0$. Thus the velocity  of
the fluid in the jet is given by
$$
\frac 12 mv^2= \mu_1-\mu_0,
\eqno (1.12)
$$
a familiar classical result.
We find then that
$$
-\hbar\frac {d\phantom t}{dt}(\theta_{jet}-\theta_{ambient})=\frac 12 mv^2.
\eqno (1.13)
$$
This accumulating phase slip must be accounted for by the passage of
vortices in the shear flow bounding the jet [11].

To verify this we note that the shear flow has vorticity $v$ per unit
length. Since $\theta$ winds through $2\pi$  as we encircle a vortex, the
quantum of circulation is $2\pi \hbar/m$. There are therefore $mv/2\pi
\hbar$ vortices per unit length in the jet boundary. Each of these
vortices will be convected with the velocity field due to all the {\it
other} vortices, and, by arguments familiar from the calculation of the
force on the windings of a solenoid, this velocity is $v/2$. Thus
smoke-ring-like vortices are shed at the orifice and convected
downstream at a rate of $\frac 12 mv^2/2\pi \hbar$ vortices per second.
Each vortex allows a phase-slip of $2\pi$, so the phase slip between
points inside the jet and those outside accumulates at a rate $\frac 12
mv^2/\hbar$, in comfortable conformity with (1.13).

This result is an example of Anderson's  general relation,  proved in
Appendix B of [11], between  the difference in the Bernoulli constant
at two points and rate of passage of vortices across any curve
connecting them.

\vskip 40pt
\line {\bf 3 Numerical Procedure and Results.\hfil}

  For the purpose of our numerical work we use the Gross-Pitaevskii
equation in the following form.
$$ i \partial_t \psi = - \bigtriangledown^2 \psi +
(|\psi|^2 - 1) \psi. \eqno(3.1) $$
 This corresponds to redefining the units for $x$ and $t$, and
the normalization of $\psi$. In these new units the speed of sound $c$
is equal to $\sqrt 2$.  As we mentioned earlier, we confine our
study to two space dimensions.

We implemented the time evolution using a stabilized leapfrog algorithm
of second order accuracy. We used a $200 \times 200$ lattice with the
lattice spacing in spatial directions chosen to be equal to 0.2 and the
spacing in time direction equal to 0.004.  Evolution was stable with
these parameters and the total density was conserved to within 1-2
percent during the evolution. The calculations were performed on SPARC
10 workstation at ITP Santa Barbara and on HP735 workstation at IOP
Bhubaneswar.

We have studied three different cases corresponding to various
geometries and sizes of the orifice. Figures 1a, 1b and 1c show the
contour plots for the initial configurations of $\psi$ corresponding to
these three cases. The order-parameter phase, $\theta$, is equal to
zero everywhere for these initial configurations. A fixed boundary
condition $\psi = 0$ is maintained at  x = 0. At the other boundary in
x direction (x = 40 in our simulations) free boundary conditions are
used, while periodic boundary conditions are used in the y direction.
The midpoint of the orifice is at x = 20. Figure 1a shows the case of
orifice with sharp edges where $\psi = 0$ is kept fixed on a line along
the y axis (at x = 20), leaving the opening for the orifice. Figure 1b
shows the case when the edges of the orifice are semicircles, and Fig.
1c shows orifice, again with rounded edges, but with a wider opening.
Dense regions of contours correspond to the healing length in which
$\psi$ grows from the value $\psi = 0$ to  the value $|\psi| = 1$.

We begin by using a relaxation method to obtain initial solutions of the G-P
equation with the boundary conditions described above.  We evolve an
initially chosen configuration using the diffusive equation
$$
\partial_t
\psi = \bigtriangledown^2 \psi - (|\psi|^2 - 1) \psi.
\eqno(3.2)
$$
After about 2000 time steps  $\psi(\bf x)$ converges to a solution of
the time independent G-P equation. These static solutions are the
configurations shown in figures 1a, 1b and 1c. They become the initial
condition for $\psi$ for the time evolution using equation (3.1).

We tried several procedures for generating a superflow.  Our first
attempt utilised a piston moving along the positive x axis
from x = 0. The piston was given a small uniform acceleration until it reached
a suitable velocity. Shedding of vortex-antivortex pairs was observed
using this method. Unfortunately, because of the finite size lattice, a
constant piston velocity led to an increasing average velocity for the
superflow through the orifice.  This method, therefore, was not very
suitable in determining the final asymptotic velocity of the superfluid
through the orifice.

  The second method  consisted of the addition of a source term  of the
  form $i w \psi/|\psi|^2$ to the G-P equation. Here, $w$ is the
strength of the source which is taken to be zero initially and
increased slowly to a suitable maximum value during the course of the
simulation. The source was taken to be non-zero along the line x =
0.4.  This greater control allowed by this procedure provided a
constant average superfluid velocity. The results reported in the
following paragraphs make  use of this method.

While we used equation (3.1) in most of the region we found it
convenient to use the  diffusive equation (3.2) in the large x region.
This helped to damp out sound waves so that the vortex shedding could
be studied in detail for large time durations. For the orifice with sharp
edges, equation (3.1) was used for region between x=0 to x=30, while
equation (3.2) was used between x=30 to x=40. For orifices with rounded
edges, these regions were respectively x=0 to x=35 and x=35 to x=40.
Since the dissipative equations were used for regions sufficiently far
from the orifice, in the direction of the flow, it did not interfere
with the process of vortex shedding.

 Figures 2, 3 and 4 show the details of the process of vortex shedding
for the three types of orifices. In these plots, solid lines are used
to plot contours of $\rho$ and dashed lines are used for $\theta$
contours. As described above, bubbles are pulled off the edges of the
orifice. These form the cores of an vortex-antivortex pair which
detaches from the edge and is swept along the flow. It is interesting
to note that, for the orifice with rounded edges, vortex shedding does not
occur at the point of smallest opening, but further downstream. This
is not uncommon in boundary layer separation phenomena. It would be
interesting to investigate what factors determine the point of vortex
shedding for such cases.

Figures 5a, 5b, and 5c show the plots of the average velocity of the
fluid flow through the different orifices as a  function of time. We
see that the flow increases uniformly, reaching, on the average, a
critical velocity asymptotically. This
critical velocity is roughly  $0.4$ (recall $c=1.414$) for all three
cases, irrespective of the shape of the edges of the orifice, or the
size of the opening.  We see that the fluid velocity first increases,
and then there is a sudden decrease in the velocity. Looking at the
contour plots at these times we see that the sharp decrease  in the
velocity  coincides with the shedding of a vortex-antivortex pair.
Actually, for rounded edges (figures 4b and 4c), the vortex shedding
slightly precedes the drop in the fluid velocity. This is not
unexpected as the vortex is produced downstream from the midpoint of
the orifice, while the fluid velocity is calculated at the midpoint. It
will take a while before the vortex shedding can  affect the midpoint
velocity.  This further supports the idea that the drop in fluid
velocity is caused by vortex shedding.  Each successive large jump in the
fluid velocity corresponds to a vortex-antivortex pair being shed.

 An intuitive understanding of this sudden jump in the fluid velocity
can be given in the following way.  As the core of the vortex buldges
out of the edge, it effectively narrows the orifice. This leads to an
increase in the fluid velocity through the orifice which in turn
further lowers the pressure. Eventually the vortex is shed and the
backwards fluid motion around the vortex tends to reduce the net  flow
through the orifice. A significant part of the kinetic energy of
the flow is transferred to the vortex.

\vskip 40pt
\line {\bf 4. Discussion \hfil }

The  vortex shedding process exhibited in this paper is probably not directly
relevant for interpreting the  experiments. The critical velocities
reported in refs. [4,5] seem too low to cause cavitation. It is more
likely that some form of quantum tunneling pre-empts the classical
mechanism. It is, however, well worth obtaining a better understanding of
the classical mechanism. The spontaneous shedding of vortices seen in
our computation demonstrates  that, at least for G-P fluids in two
dimensions, the tunneling  barrier  to creating a vortex disappears
above some critical velocity. This critical velocity needs to be estimated
if the purely classical effect is to be distinguishable from the
more interesting macroscopic quantum tunneling.

How does the vortex tunnel? Homogeneous quantum nucleation [12,13] is
not possible when there is no normal fluid to select a preferred
reference frame, but there are several competing channels  that use
the boundary to break galilean invariance.  One possibility is to
create a complete vortex ring surrounding the orifice. This, however,
requires a large action. A more popular scenario is the quantum
nucleation of a semicircular segment of vortex terminating on the
boundary [14,15,16,17].  This segment is then swept into the flow where
it grows by extracting energy from the bulk motion.  Because this
likely scenario breaks the axial symmetry, two-dimensional computations
are not adequate for determining the classical critical velocity  ---
if it exists. (Both the thin vortex models of [14,15] and the static
G-P analysis in [16] suggest that a tunneling barrier persists even for
large flow velocities.) It is therefore essential to perform a full
three-dimensional time-dependent computation for an orifice
with a suitable asperity on which the vortex can form.

\vskip 40pt
\line{\bf Acknowledgements\hfil}

Part of this work was carried out at the ITP Santa Barbara and we would
like to the thank the staff and members of the ITP for their
hospitality.  We were supported by the National Science Foundation
under grant numbers PHY89-04035 (ITP) and by DMR91-22385 and its
successor DMR94-24511 (University of Illinois).  We would also like to
thank Nigel Goldenfeld for alerting us to the existence of  refs
[8,9].

\vskip 40pt
\line{\bf References\hfil}

\item{[1]} For example see: G.~K.~Batchelor {\it An
Introduction to Fluid Mechanics.} (Cambridge University Press 167)
Chapter 5.

\item {[2]}  W.~Zimmermann  Jr.; O.~Avenel,
E.~Varoquaux, Physica B {\bf 165} (1990) 749;  E.~Varoquaux, O.~Avenel,
Physica B {\bf 197} (1994) 306.

\item{[3]} A.~Amar, Y.~Sasaki, R.~L.~ Lozes, J.~C.~Davis, R.~E.~Packard, Phys.
Rev. Lett. {\bf 68} (1992) 2624.

\item{[4]} J.~C.~Davis, J.~Steinhauer, K.~ Schwab,
Yu.~M.~Mukharsky, A.~Amar, Y.~Sasaki, R.~E.~ Packard, Phys. Rev Lett.
{\bf 69} (1992) 323.

 \item{[5]} G.~G.~Ihas, O.~Avenel, R.~Aarts,
R.~Salmelin, E.~Varoquaux, Phys. Rev. Lett. {\bf 69} (1992) 327.

\item{[6]}  L.~P.~Pitaevskii,  Zh. Eksp. Teor. Fiz. {\bf 40}
(1961) 646, Translated in Sov. Phys. JETP {\bf 13} (1961) 451.

\item{[7]} E.~P.~Gross, Nuovo Cimento {\bf
20} (1961) 454; J. Math Phys. {\bf 4} (1963) 195.

\item{[8]} T.~Frisch, Y.~Pomeau, S.~Rica,
    Phys. Rev. Lett. {\bf 69} (1992) 1644.

\item{[9]}  Y.~Pomeau, S.~Rica, Phys. Rev. Lett. {\bf 71} (1992)
247.

\item{[10]} E.~Madelung, Z. Phys {\bf 40} (1927) 322.

\item{[11]} P.~W.~Anderson.  Rev. Mod. Phys. {\bf 38} (1966) 298.

\item{[12]} J.~S.~Langer, M.~E.~Fisher, Phys. Rev. Lett. {\bf 19} (1967) 560.

\item{[13]} J.~S.~Langer J.~D.~Reppy, Progress in Low Temperature Physics {\bf
6} (1970) 1.

\item{[14]} G. Volovik, Pis'ma, Zh. Eksp. Teor Fiz {\bf 15} (1972) 116 (JETP
Letters {\bf 15} (1972) 81).

\item{[15]} C.~M.~Muirhead, W.~F.~Vinen, R.~J.~Donnelly, Phil Trans. Roy. Soc.
{\bf A311} (1984) 433.

\item{[16]} M.~Bernard, S.~Burkhart, O.~Avenel, E.~Varoquaux, Physica B {\bf
194-196} (1994) 499; Phys. Rev. Lett. {\bf 72} (1994) 380.

\item{[17]} K.~W.~Schwarz, Physica B {\bf 197} (1994) 324; Jour. Low
Temp. Phys. {\bf 93} (1993) 1019; Phys. Rev. Lett. {\bf 71} (1993) 259.

\vfil\eject

\vskip 40pt
\line{\bf Figure Captions.\hfil}

\item{Fig 1.} Initial configurations. The solid lines are contours of density.

\item{Fig 2.} Early stage of flow. The solid lines are contours of density.
Dashed lines are contours of order-parameter phase. The heavy dashed lines
are branch cuts where the order-parameter phase jumps through $2\pi$.

\item{Fig 3.} Same as Fig 2., but at later time.

\item{Fig 4.} Same as Fig 3., but at later time.

\item{Fig 5.} Flow at midpoint of aperture. Each sharp dip in the velocity
corresponds to a vortex-antivortex pair being shed.

 
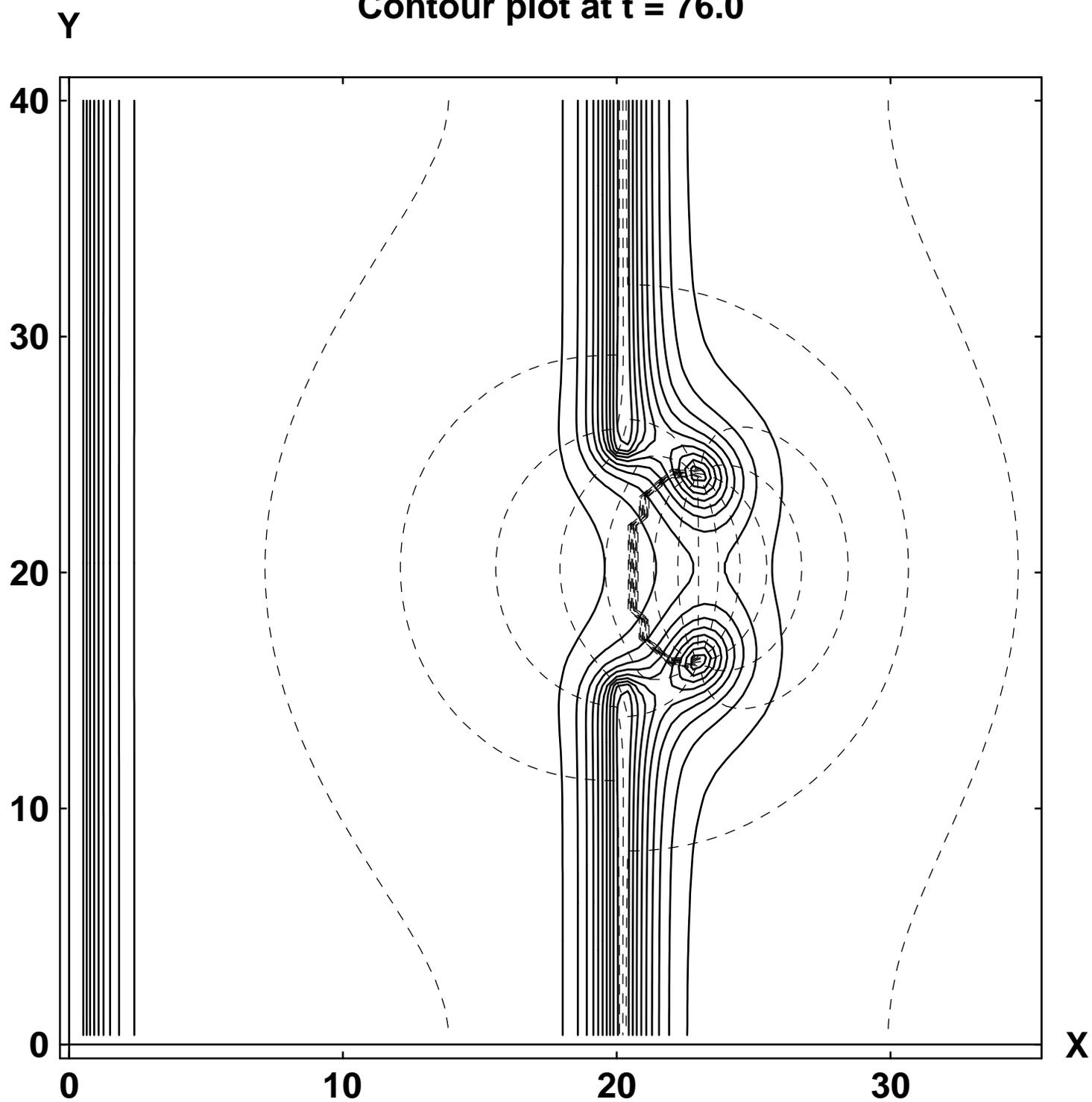
\bye